\documentclass[aps,twocolumn,longbibliography]{revtex4-1}

\usepackage{bm}
\usepackage{graphicx}
\usepackage{amsmath}
\usepackage{amssymb}
\usepackage{amsfonts}
\usepackage{float}
\usepackage{color}
\usepackage[utf8x]{inputenc}

\begin{document}

\title{Hysteresis, neural avalanches and critical behaviour near a first-order transition of a spiking neural network}

\author{Silvia Scarpetta$^{1,2}$}
\author{Ilenia Apicella$^{3}$}
\author{Ludovico Minati$^{4}$}
\author{Antonio de Candia$^{2,5}$}

\affiliation{$^1$Dipartimento di Fisica ``E. Caianiello'', Universit\`a di Salerno, Fisciano (SA), Italy}
\affiliation{$^2$INFN, Sezione di Napoli, Gruppo collegato di Salerno, Fisciano (SA), Italy}
\affiliation{$^3$Dipartimento di Fisica e Astronomia ``G. Galilei'', Universit\`a di Padova, Italy}
\affiliation{$^4$Complex Systems Theory Department, Institute of Nuclear Physics Polish Academy of Sciences (IFJ-PAN), Kraków, Poland}
\affiliation{$^5$Dipartimento di Fisica ``E. Pancini'', Universit\`a di Napoli Federico II,\\
Complesso Universitario di Monte Sant'Angelo, via Cintia, 80126 Napoli, Italy}

\begin{abstract}
Many experimental results, both in-vivo and in-vitro, support the idea that the brain cortex operates near a critical point, and at the same time works as a reservoir
of precise spatio-temporal patterns. However the mechanism at the basis of these observations is still not clear. In this paper we introduce a model
which combines both these features, showing that scale-free avalanches are the signature of a system posed
 near the spinodal line of 
a first order transition,
with many spatio-temporal patterns stored as dynamical metastable attractors.
Specifically, we studied a network of leaky integrate and fire neurons, whose connections are the result of the learning of multiple
spatio-temporal dynamical patterns, each with a randomly chosen ordering of the neurons.
We found that the network shows a first order transition between
a low spiking rate disordered state (down), and a high rate state characterized by the emergence of collective activity
and the replay of one of the stored patterns (up).
The transition is characterized by hysteresis, or alternation of up and down states, depending on the lifetime of the metastable states.
In both cases, critical features and neural avalanches are observed. 
Notably, critical phenomena occur at the edge of a discontinuous phase transition, as recently observed in a network of glow lamps.
\end{abstract}

\maketitle

\section{Introduction}

Recently, many experimental results have supported the idea that the brain operates near a critical point
  \cite{Plenz2003,in-vivo,pasquale,shriki,chialvo2010,plenzlibro,specialissue,viola},
as reflected by power law distributions of avalanche sizes and durations.
The maximization of fluctuations near a critical point is believed to play an important role in the ability of the brain to respond to a wide range of inputs,
to process the information in an optimal way \cite{copelli-opt,shew-opt,shew-2,shew-3,shriki2},
and to enhance stimulus discriminability \cite{tomen-opt}.
The theoretical framework commonly used to explain this behaviour is the branching process, which undergoes a second order transition
when the branching parameter becomes greater than one. The order parameter, that is the probability to observe an infinite avalanche,
indeed continously grows above the transition.

On the other hand, metastability and hysteresis are ubiquitous in the brain.
They are related to the ability of the brain
to  sustain stimulus-selective persistent activity for working memory \cite{Sejnowsky}.
The brain rapidly switches from one state to another in response to a
stimulus, and it may remain in the same state for a long time after the end
of the stimulus, suggesting the existence of a repertoire of metastable states.
The presence of metastability and criticality could be reconciled if the system is posed near the edge of instability (spinodal line) of a first order transition.

Recently it has been shown that a simple network of glow lamps (nonlinear devices that share some similarity with leaky neurons)
show a critical behaviour near the edge of a first-order (discontinuous) phase transition \cite{lamps}. 
Critical phenomena and avalanches indeed
emerge not only in second order transitions, but also in discontinuous ones, as one enters the metastability region
and approaches the spinodal curve \cite{spinodal3,spinodal3bis}.
Close to the spinodal, which for long range interactions
denotes the limit of existence of the metastability region,
transition precursors are observed which follow power-law
scaling having a cut-off diverging to infinity on the spinodal
itself;
examples are found, for instance, in geophysical
phenomena, breakdown of solids, and spontaneous network
recovery \cite{spinodal,spinodal2,spinodal4,spinodal5}.
Bistability with critical features is observed also in non-equilibrium phase transitions \cite{jordi}.

In the present paper, our goal is to understand if a first order transition with spinodal instabilities may be a correct scenario in neural cortical experiments.
We study a simple stochastic leaky spiking model, whose quenched disordered connectivity is the result of learning multiple spatio-temporal patterns,
and simulate the spontaneous activity of the network applying a Poissonian noise to individual neurons, 
related to the spontaneous neuro-transmitter release at individual synapses,
as well as other sources of inhomogeneity and randomness that determine an irregular background synaptic noise.

We observe that  there is a parameter region characterized by a first %
order transition which notably shows hysteresis and metastability.
The phase transition is between a low activity state, with uncorrelated firing and low rate, and a state characterized by collective activity
with high firing rate and high spatio-temporal order,
where one of the stored patterns emerges. %
At higher values of the noise,  or smaller network sizes,
lifetimes of the states become smaller then the observation time, so that instead of hysteresis we observe an alternation of the two phases.

Scale-invariant
spatio-temporal avalanches occur at the edge of the transition, both inside the hysteresis region
(lifetimes of the metastable states longer than the observation time)
and near the alternating region (lifetimes smaller than the observation time).
Notably we find that the
average avalanche size as a function of the avalanche duration $s(T)$, collapse on a universal power law 
with an exponent
close to the experimental one \cite{naturePhys2015,PRLfriedmann}.

Another important characteristic of avalanches in the brain, is that they contain highly repeatable patterns, both in-vitro \cite{PlenzBeggs2004} and in-vivo \cite{copelli2016},
supporting the hypothesis that scale-free neural avalanches are the signature of a critical behaviour in a system that has stored multiple dynamical spatio-temporal patterns.
Notably, it has been shown \cite{copelli2016}  that spike
avalanches, recorded from freely-behaving rats, form repertoires that emerge in waking, recur during sleep, are diversified by
novelty and contribute to object representation. They constitute distinct
families of recursive spatio-temporal patterns, and a significant number of those patterns
were specific to a behavioral state.

Storing precise spatio-temporal patterns as dynamical attractors of the network is a useful strategy for brain functioning,
coding and memory, and many experimental results on the replay of precise spatio-temporal patterns of spikes suggest this possibility
\cite{plos16,plos18,plos20,plos21,plos25,plos27}.

Our model captures such additional features of neuronal avalanches,
such as the underlying first-order transition between attractor dynamics and quiescence, 
the stable recurrence of particular spatio-temporal patterns,
and the conditions under which these precise and diverse patterns can be retrieved.

Critical avalanches were observed in a leaky integrate and fire model of neurons \cite{plos,frontiers},
but for a single value of the noise and of the size of the system, where no hysteresis was observed,
and the type of underlying phase transition was not
thoroughly investigated. 
The role of first order phase transition for criticality in cortical networks was firstly pointed out by Ref.\ \cite{levinaPRL2009},
and successively elaborated in a leaky integrate and fire model \cite{millman}. However, as shown in \cite{maritan}, in such models
criticality emerges only with a definition of avalanches that takes in account the causality of different firings. To our knowledge, our model is the first
that exhibits neural avalanches at the edge of a first order transition, and are identified with the same temporal proximity criterion used in experiments.

\section{Results}

We study a model of leaky integrate and fire neurons, whose connectivity is the result of the learning of multiple spatio-temporal patterns,
using a learning rule inspired by spike-time dependent plasticity (STDP).
The emerging spontaneous dynamics is simulated in presence of noise, 
with fixed sparse connections, and a small fraction of leader neurons (see Appendix).
Two parameters characterize the dynamics. The first is the parameter $H_0$ that sets the average strength of the connections,
the second is the parameter $\alpha$ that is  the coupling of each neuron to the noise.
The number of neurons goes from $N=3000$ to $N=12000$, with a number of encoded patterns from $P=2$ to $P=10$.

We simulate the spontaneous dynamics of the model in absence of external stimuli, as a function of the parameters $H_0$ and $\alpha$.
Depending on the value of the parameters,
two different  dynamical states are distinguishable:
a quiescence state (``down'' state), characterized by uncorrelated spiking with low firing rate, and an active state (``up'' state)
characterized by a high rate and high
spatio-temporal order, and by a long-lasting collective replay of stored patterns.

\begin{figure}[tbp]
\begin{center}
A \includegraphics[width=5cm]{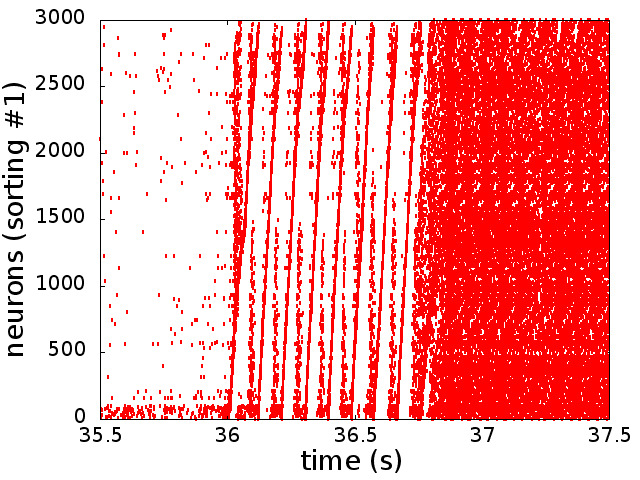}
\\
B \includegraphics[width=5cm]{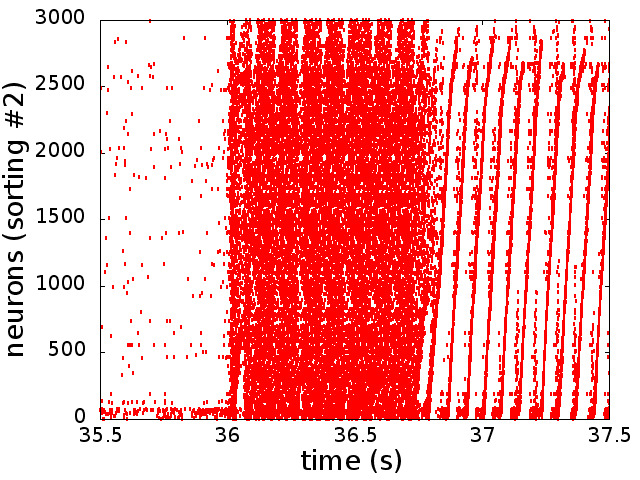}
\\
C \includegraphics[width=5cm]{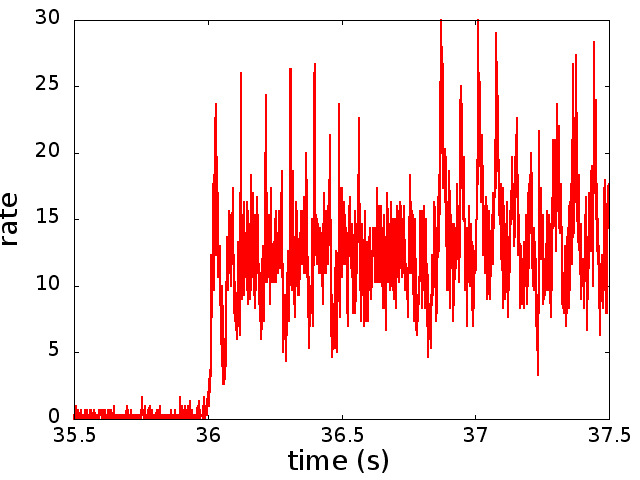}
\\
D\includegraphics[width=5cm]{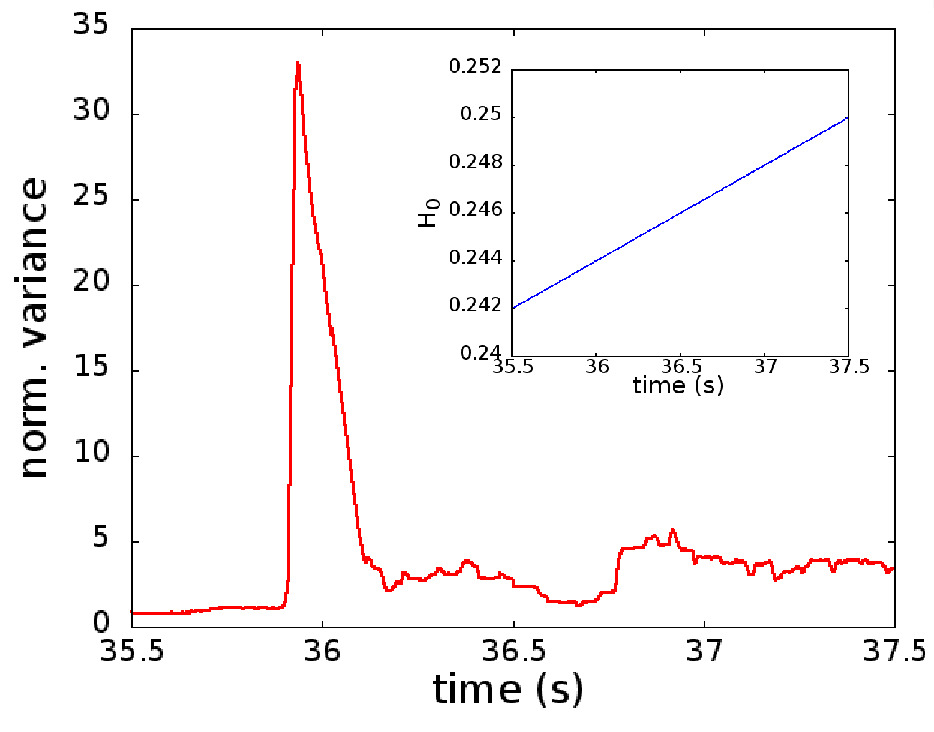}
\caption{%
Spontaneous dynamics for $N=3000$ at noise $\alpha=0.033$, showing a transition while we increase parameter $H_0$ from $0.1$ to $0.3$ in 50 seconds.
A transition, between a quiescence state (down) and a  state with emerging of collective replay of stored patterns (up),
is shown to occur at time $t=36$ seconds, corresponding to a value $H_0=0.244$.
Raster plots (A) and (B) show the same spontaneous spiking activity with two different sorting of neurons on the vertical axis.
In (A) neurons are sorted by the spiking time in pattern 1, while in (B) they are sorted by the spiking time in pattern 2.
The replay of the pattern corresponding to the sorting on the vertical axis is apparent by the saw-tooth ordering of the spikes:
from time $t=36$ s to time $36.75$ s stored pattern number two is replayed, while from  $t=36.75$ s stored pattern number one is replayed.
Note that, when a pattern is replayed, there seems to be a lower density of dots, due to the fact that the dots overlap.
In (C) we show the average instantaneous firing rate, corresonding to both the raster plots A and B.
The rate is measured as the total number of spikes in a time interval of $\Delta=1$ ms divided by $N\Delta$.
In (D) we show the normalized variance of the firing rate (see text).
In the down state normalized variance is equal to one, while in the up state it goes to values $\sigma>3$, signaling a non-Poissonian dynamics, with temporal clustering.
When the system has a transition to the up state, the firing rate abruptly increases, and normalized variance has a peak.
Inset: value of $H_0$ as a function of time.
}
\label{fig:1}
\end{center}
\end{figure}

To characterize the dynamics, we define the instantaneous rate $r$ and the normalized variance $F$ (also called Fano Factor or index of dispersion), as follows:
\begin{subequations}
\begin{align}
r&=\frac{N_{\text{tot}}}{N\Delta},
\\
F&=N\Delta\frac{\langle r^2\rangle-\langle r\rangle^2}{\langle r\rangle},
\label{eqnormvar}
\end{align}
\end{subequations}
where $N_{\text{tot}}$ is the total number of spikes over all the network
in the time interval $\Delta$, $N$ is the number of neurons of the network, and the average $\langle\cdots\rangle$
is evaluated over a sliding window $[t-T,t+T]$. We use a time interval $\Delta=1$ ms to compute the firing rate, and
a half-width $T=100$ ms for the sliding window.
The normalized variance (\ref{eqnormvar}) can also be written as
\begin{equation}
F=\frac{\langle N_{\text{tot}}^2\rangle-\langle N_{\text{tot}}\rangle^2}{\langle N_{\text{tot}}\rangle},
\end{equation}
showing that, if neurons are uncorrelated and Poissonian, then $F=1$.
If $F>1$ the spiking activity is over-dispersed: this corresponds to the existence of clustered activity,
with some intervals having a much higher activity than the mean, and others a very low activity, compared to a Poisson distribution.
If on the other hand $F<1$, activity is under-dispersed, with many intervals having spike counts close to the mean.

In Fig.\ \ref{fig:1}, we show the dynamics of the network at fixed noise $\alpha=0.033$, while we increase parameter $H_0$ from $0.1$ to $0.3$ in 50 seconds.
(Note that, throughout the paper, the time is always measured as the ``physical'' time appearing in Eq.\ (\ref{eqSRM}), not the CPU time needed to simulate the system.)
At low values of $H_0$, the spiking rate is low, less than 1 Hz, and normalized variance is near to one, signaling uncorrelated Poissonian activity (down state).
At time $t=36$ seconds, when $H_0$ reaches the value $H_0=0.244$, we observe an abrupt transition 
to a state corresponding to the sustained collective replay of
one of the stored patterns with high firing rate (up state).
In Fig.\ \ref{fig:1}A and B we show the raster plots of the dynamics in the same interval of time, with neurons ordered on the vertical axis by the two patterns encoded in the
network. It can be seen that the transition to the up state corresponds to the replay of one of the stored patterns. 

Note that Fig.\ \ref{fig:1}A and B refer to the same spike train. The different ordering on the vertical axis makes the spikes appear in a saw-tooth shape when the pattern
corresponding to the order is replayed, while they appear as completely random (and deceitfully denser) when another pattern (not corresponding to the ordering of the vertical axis)
is replayed. In Fig.\ \ref{fig:1}C we show the rate corresponding to the dynamics shown in both the raster plots \ref{fig:1}A and B. It can be seen that, at time $t=36$ seconds,
when the collective replay of the first patterns starts, the rate sharply increases from a very low value, to an average value of 13 Hz, fluctuating between 5 and 30 Hz.
Correspondingly, normalized variance $F$ jumps from one (Poissonian dynamics) to a value between 2 and 5 (temporally clustered).
Note that values of normalized variance greater than one, are found experimentally in persistent activity in cortical circuits \cite{barbieri}.
Exactly at the transition, the normalized variance has a high peak.

\subsection{Hysteresis and first order transition}

\newcommand{\downtoup}{\mbox{{\em down} $\to$ {\em up}} }
\newcommand{\uptodown}{\mbox{{\em up} $\to$ {\em down}} }

\begin{figure}[tbp]
\begin{center}
A \includegraphics[width=5cm]{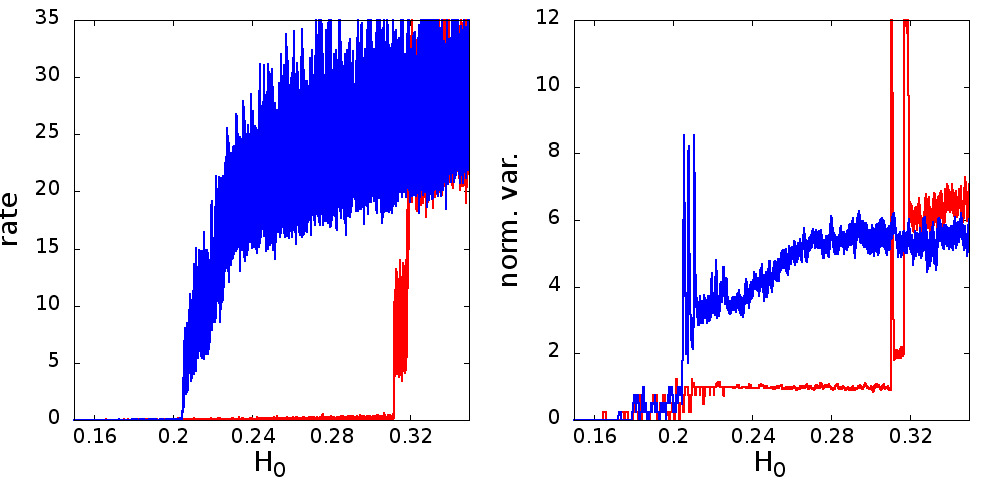}
\\
B \includegraphics[width=5cm]{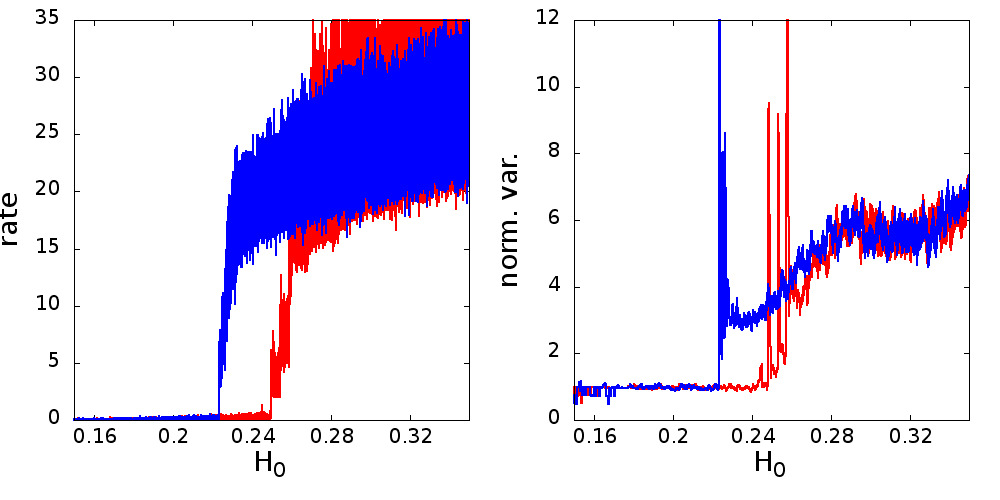}
\\
C \includegraphics[width=5cm]{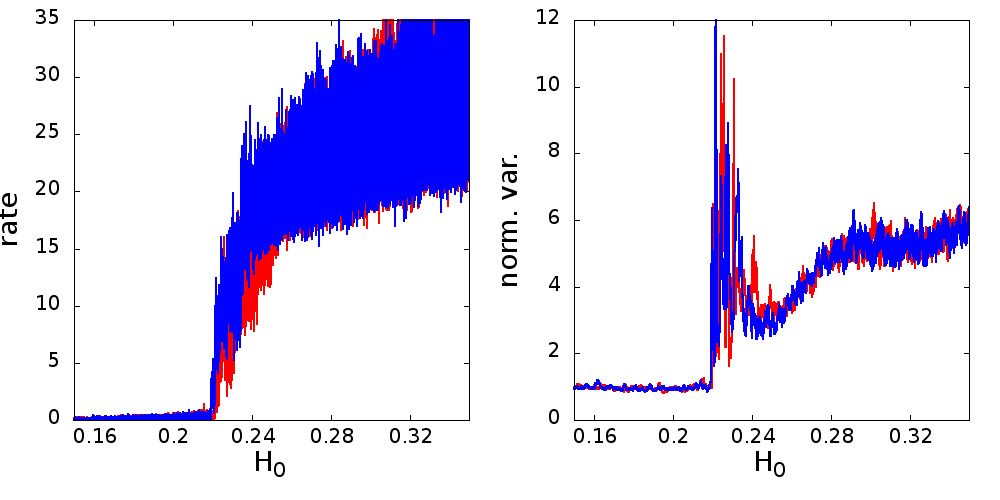}
\\
D \includegraphics[width=5cm]{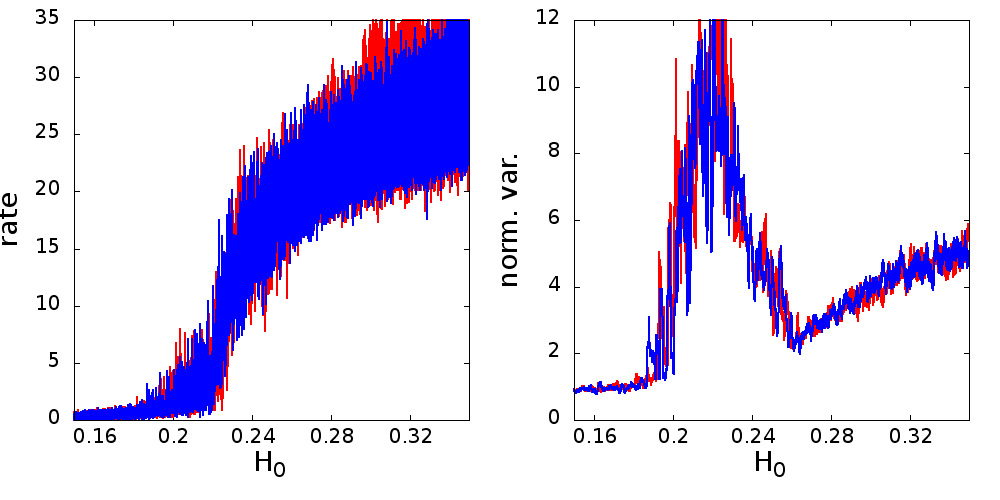}
\end{center}
\caption{Firing rate and normalized variance during spontaneous dynamics while sweeping $H_0$ at fixed $\alpha$, for $N=3000$ and $P=2$, and for values of the noise
A) $\alpha=0.015$, B) $\alpha=0.03$, C) $\alpha=0.045$, D) $\alpha=0.1$, showing a hysteretic behaviour at low values of the noise.
The strength of connections $H_0$ is increased from $H_0=0.1$  to $H_0=0.4$
during the first 50 seconds of the simulation (red line), and then decreased back to $H_0=0.1$ during the last 50 seconds (blue line),
with a linear schedule.
Transitions between two dynamical states, a ``down'' state with low rate and normalized variance equal to one,
and an ``up state'', with much higher rate and normalized variance $F>3$, are observed at different values of $H_0$ while ramping up or down,
showing hysteresis at low values of the noise.
Peaks in the normalized variance signal the transitions. At high values of the noise, $\alpha\ge 0.045$, there is an interval of $H_0$, around $H_0=0.22\sim 0.24$,
where multiple transitions \downtoup and \uptodown are observed.
}
\label{fig:hysteresis}
\end{figure}

The observed discontinuous behaviour of the rate and variance suggests that the transition is of a first order kind. An important characteristic of first order transitions
is hysteresis, so here we investigate if our model actually shows hysteresis while varying parameters $H_0$ and $\alpha$.
At fixed value of $\alpha$, we start with the system in the down state and $H_0=0.1$, and cycle $H_0$ from 0.1 to 0.4 in the first 50 seconds, and back from 0.4 to 0.1 in
the last 50 seconds. In Fig.\ \ref{fig:hysteresis} we show the instantaneous rate and variance as a function of $H_0$ in the first half of the run (increasing $H_0$, red lines)
and in the second half (decreasing $H_0$, blue lines). Both rate and variance are averaged over four different runs, with different realization of stochastic noise.

\begin{figure}[tbp]
\begin{center}
\includegraphics[width=5cm]{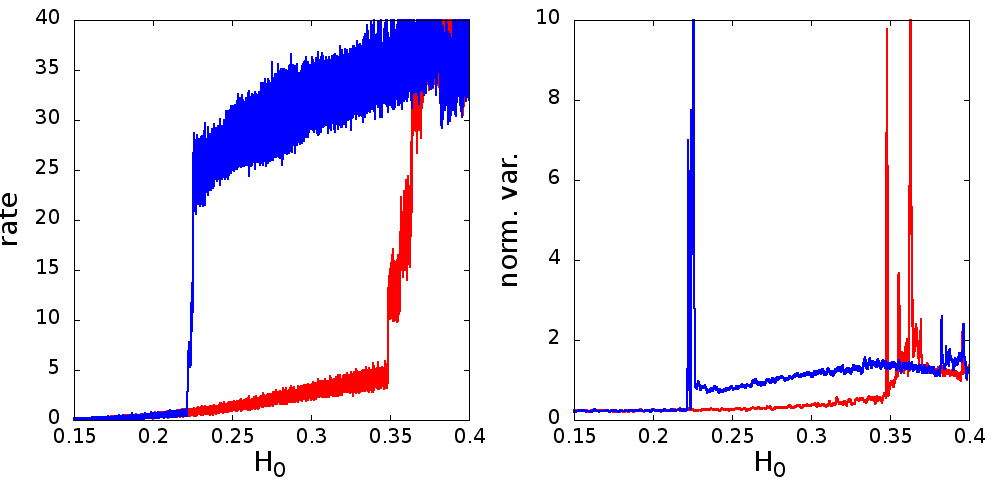}
\end{center}
\caption{Firing rate and normalized variance while sweeping $H_0$ as in Fig.\ \ref{fig:hysteresis}, for $N=12000$ and $P=10$, and noise $\alpha=0.01$.
The hysteretic behaviour is robust with respect to increasing size and number of patterns.
}
\label{fig:hysteresisbis}
\end{figure}

For low values of the noise, we observe a strong hysteretic behaviour of the dynamics.
Looking for example at figure \ref{fig:hysteresis}A,
where  $\alpha=0.015$, we observe that when $H_0$ is increased, \downtoup transitions take place between $H_0=0.31$ and 0.32. As the rate and variance are averaged over four runs,
there are actually four different transitions at slightly different values of $H_0$,
depending on the realization of the stochastic noise.

On the other hand, when $H_0$ is decreased,
\uptodown transitions take place at lower values of $H_0$, in this case around $H_0=0.2$.
In \ref{fig:hysteresis}B, C, and D, we show the same experiment for higher values of the noise parameter $\alpha$. When $\alpha\ge 0.45$, the value of the rate and variance does not
depend anymore on the history, and is equal within fluctuations when $H_0$ is increased or decreased.
Moreover one can observe multiple back and forth transitions \uptodown and \downtoup, during the same run, giving rise to a large peak in the variance.
In Fig.\ \ref{fig:hysteresisbis}, we show the behaviour of the system for a higher value of the number of neurons, and of the number of patterns.

Hysteresis is a hallmark of first order transitions, characterized by the presence of two (or more) possible states of the system, separated by barriers difficult to overcome.
If the systems stays in one state, it will tend to remain in that state also when external parameters would favour another one. Therefore the state of the system, depends
on the past history, for example if $H_0$ is being increased or decreased.
The nucleation time, i.e. the lifetime of metastable states, depend critically on the range of the connections. If
the model is characterized by long range connections, one could expect a ``mean field like'' behaviour, with the transitions from the metastable to stable states happening
on the spinodal lines. However, in our case,
albeit the connections do not depend on distance (that is they are long range),
the number of units is not very large, so we expect that, at any point in the space of parameters, there will be
a nucleation time sufficient to switch the system from one state to the other, that can also be interpreted as a typical lifetime of the state.
Transitions \downtoup will be observed when the lifetime of the down state becomes comparable or smaller than the experimental time, taken as the inverse rate of change of $H_0$,
while on the contrary transitions \uptodown will be observed when the lifetime of the up state becomes smaller than the experimental time.
At high values of the noise, or at small system sizes, the lifetime of both up and down states becomes smaller than the experimental time, so that an alternation of up and down states can be observed.

\begin{figure}[tbp]
\begin{center}
\includegraphics[width=5cm]{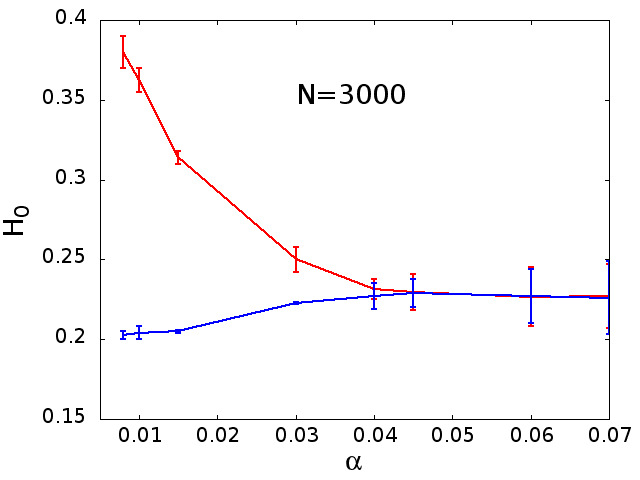}
\\
\includegraphics[width=5cm]{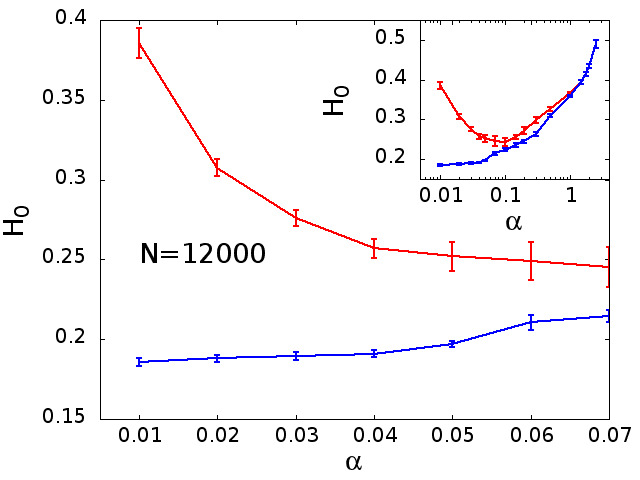}
\end{center}
\caption{(A) Hysteresis region and ``{\em pseudo-spinodal}'' lines for $N=3000$ and $P=2$.
The red line marks the (average) value of $H_0$ where the dynamics switches from down to up state when $H_0$ increases, at a fixed value of $\alpha$.
The blue line the (average) value where the dynamics switches from up to down state when $H_0$ decreases. 
Bars indicate the lowest and highest values of $H_0$ where transition was observed  (with several different realizations of the
  stochastic noise).
At high value of the noise, $\alpha>0.04$, 
we observe no hysteresis, and
bars indicate the interval in which multiple \downtoup and \uptodown transitions are observed.
(B) Same plot for $N=12000$ and $P=2$. Lifetimes of the states increase with respect to the $N=3000$ case, so that 
the up and down alternating region at these values of the noise disappears, and the hysteresis region broadens. In the inset a larger range of parameters is investigated,
showing that ``{\em pseudo-spinodal}'' lines merge at a higher value of noise and $H_0$.
}
\label{fig:phasespace}
\end{figure}

In Fig.\ \ref{fig:phasespace}, the phase space of the system for $N=3000$ (A) and $N=12000$ (B) is shown.
A red line marks the boundary where the dynamics switches from down to the up state when $H_0$ increases, at a fixed value of $\alpha$.
while a blue line marks the boundary from up state to down state when $H_0$ decreases.
Bars indicate the width of the region where the transition may happen,
namely the lowest and highest values of $H_0$ where the transition was observed,
for several realizations of the patterns and of the stochastic noise.
Inside the strip defined by the bars, one may observe multiple back and forth transitions, i.e. an alternation of down and up states.

Red and blue lines can be interpreted as ``{\em pseudo-spinodal}'' lines, that mark the point where the lifetime of the state (or nucleation time)
becomes smaller than the observation time.
While in systems with short range connections the nucleation time is independent from the size of the system, when connections are long range, as in our case,
we expect that the nucleation times grow with the size of the system. Indeed, as shown in Fig.\ \ref{fig:phasespace}B,
increasing the number of neurons from $N=3000$ to $N=12000$, the hysteresis region broadens, showing that lifetime of the states increases.

This means that the convergence of the ``pseudo-spinodal'' lines at $\alpha=0.045$, for $N=3000$, is actually a finite size effect, but the transition
is still first order at these values of the parameters.
As shown in the inset of Fig.\ \ref{fig:phasespace}B, for $N=12000$ lines meet at a much higher value of the noise, and a higher value of $H_0$.
It is reasonable to expect that, in the thermodynamic limit $N\to\infty$, the point where line meet will tend to a definite value of $\alpha$ and $H_0$,
corresponding to a second order transition point, terminating the first order transition line.

\begin{figure}[tbp]
\begin{center}
\includegraphics[width=5cm]{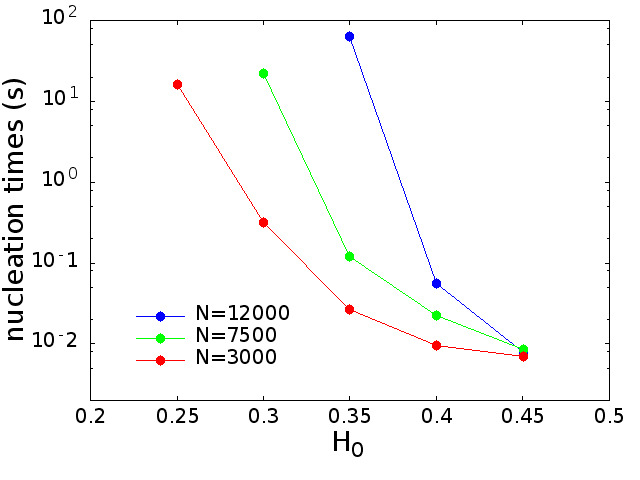}
\end{center}
\caption{
  Average nucleation times are shown for network sizes $N=12000$, $7500$, $3000$, at a loading parameter $P/N=1/1500$
and noise $\alpha=0.03$, as a function of $H_0$. Notably the nucleation time grows with system syze,
as expected in a system with long range connections undergoing a first order transition.} 
\label{fig-nucleazione}
\end{figure}

To check the behaviour of the nucleation time with network size, in Fig.\ \ref{fig-nucleazione}
  we investigate the nucleation time
  for network size $ N=12000,7500,3000 $ at loading parameter $P/N=1/1500$ and noise $\alpha=0.03$.
  Notably the nucleation time grows with the size of the system, supporting
  the hypotesis that metastable states have infinite lifetimes in the termodynamic limit, as is expected for a system with long range interactions undergoing a first order transition.

\subsection{Critical behaviour at the edge of instability}

The presence of metastability and hysteresis indicate that the transition is a non-equilibrium first order one.
We here show that, when one enters the metastable region from below, precursor phenomena in the form of scale-invariant spatio-temporal activity bursts can be observed,
 that are distributed following power laws both in size and in duration.

Indeed when approaching the \downtoup transition from below, before falling in the persistent up state, 
the network may have high fluctuations with transient periods of high activity.
One can observe that inside this short periods of high firing rate, at a finer level, the activity is made of a series of cascades or ``avalanches'',
separated by short drops in the rate, distributed with high diversity in spatio-temporal scale,  resulting in power-law distributions.

\begin{figure}[tbp]
\begin{center}
A \includegraphics[width=5cm]{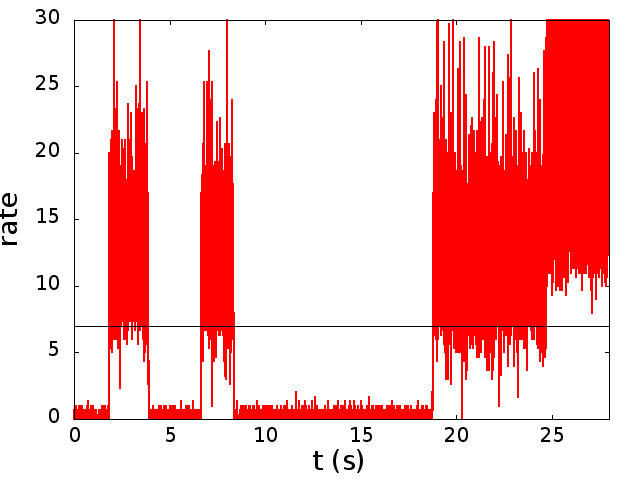}
B \includegraphics[width=5cm]{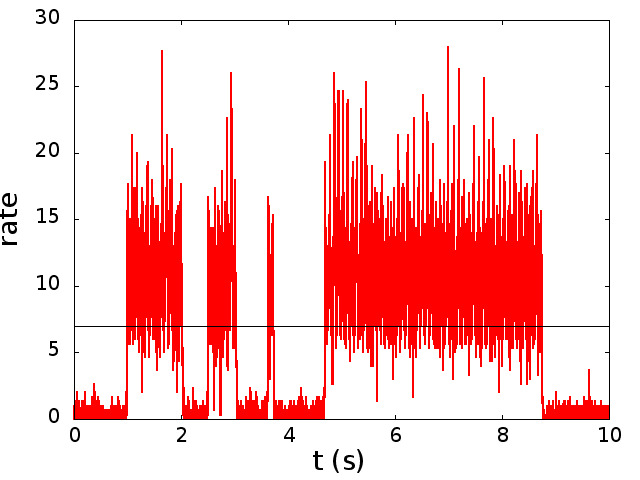}
\end{center}
\caption{Firing rate at fixed values of the noise and connection strength, for (A) $\alpha=0.03$, $H_0=0.22$ and (B) $\alpha=0.045$, $H_0=0.22$,
   at $N=3000$.
  At lower values of the noise, the system eventually falls into a state of ``persistent up''.
  We identify this as an interval of 10 seconds where the rate is always larger than 10 Hz
(last seconds of simulation in A), and stop the simulation. Avalanches are identified
  as consecutive time bins of $\Delta t=1$ ms, with a rate higher than a threshold $R_{\text{min}}=7$ Hz.
  Three or four intervals (respectively in A and B) in which the system is in an
up state are shown, that are in turn composed of many avalanches.
}
\label{fig-rate}
\end{figure}

We perform the following experiment: we fix a value of the noise $\alpha$ and connection strength $H_0$, and simulate the spontaneous dynamics of the network.
At low values of the noise, as the system is in a metastable state, that has a finite lifetime, after some non predictable time it will fall in the state of
persistent replay of one of the stored patterns. We identify this event by looking when the average firing rate of the neurons stays above 10 Hz for an interval of time longer
than 10 seconds. When the network falls in this ``persistent up'' state, we terminate the simulation and start the dynamics again from the beginning with a different realization of the noise.
During the run before falling in the state of persistent replay, we measure the rate of the network and identify the bursts of activity or avalanches.
In Fig.\ \ref{fig-rate}A, we show the rate during a run with $\alpha=0.03$, $N=3000$ and $H_0=0.22$. Note that in the last seconds
of the simulation the rate remained above 10 Hz for  10 seconds,
so the run was terminated. In the first 25 seconds, three bursts of activity can be seen, which were identified as a series of avalanches.

In contrast, for higher values of the noise,  $\alpha\ge 0.045$
for $N=3000$, 
the lifetime of the metastable states become smaller, 
and one observes an interval of values of $H_0$ where the system shows bursts of activity,
with short up and down alternation, 
without ever falling into the state of persistent
replay, as shown in Fig.\ \ref{fig-rate}B.
Indeed, in this model with structured connectivity and replay of stored patterns, the noise has a two-fold effect, on one hand it stimulates the start of a burst of activity,
i.e. initiates a short collective replay of one of the stored patterns,
but, on the other hand, it can also stop its propagation and therefore hinder its persistent replay.

Note that up and down alternation, with bursts of generalized spiking that last for many seconds, have been observed to occur spontaneously in a variety of systems and conditions,
both {\em in vitro} \cite{cossart,shu} and {\em in vivo} \cite{petersen,luczak}.
These bursts are composed by many avalanches.

As recently pointed out in Ref.\ \cite{maritan}, two different methods have been used to define avalanches. The first is based on the temporal proximity of neural activity,
so that if activity happens in contiguous time bins, it is considered belonging to the same avalanche. The second takes in account the causality of firing, so that activity of two neurons
belong to the same avalanche if the spike of the first neuron directly causes the second neuron to fire.
A novel tool for detect cascades of causally-related events experimentally has been found, and it shows that
indeed neuronal avalanches are not merely composed of causally-related events \cite{beggs17}.
We define avalanches according to first method, that is the one used in experiments, where causal information is not usually accessible.
In particular, we use the methods
implemented by Refs.\ \cite{ava1,ava2,shew-opt},
which altered the original method used by Ref.\ \cite{Plenz2003}, to make it more suitable when the activity of a large number of neurons is measured.
In Ref.\ \cite{ava2}, both methods were used, finding consistent results.
Avalanches are therefore defined as periods of time where the population firing rate exceeds a threshold.
As the population firing rate distribution is bimodal, reflecting the existence of the two phases,
  we set the threshold slightly higher then the minimum of the bimodal rate
  distribution, to minimize the probability of concatenating different avalanches.
  The minimum slightly changes with system size, therefore
  we use a threshold of $R_{\text{min}}=7$ Hz at $N=3000$, and  $R_{\text{min}}=10$ Hz at $N=6000$ and $12000$,
  using a time bin of $\Delta t=1$ ms to measure the population firing rate.
Note that the rate is defined in terms of average spiking rate of single neurons, therefore a rate $R$ in Hz corresponds to $RN/1000$ spikes per milliseconds,
where $N$ is the number of neurons.
We define the size of an avalanche as the total number of spikes, that is the integral of the rates over the avalanche duration.

\begin{figure}[tbp]
\begin{center}
A \includegraphics[width=5cm]{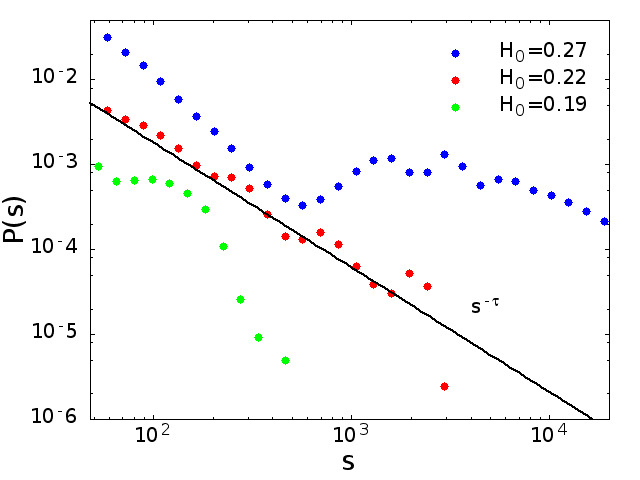}
\\
B \includegraphics[width=5cm]{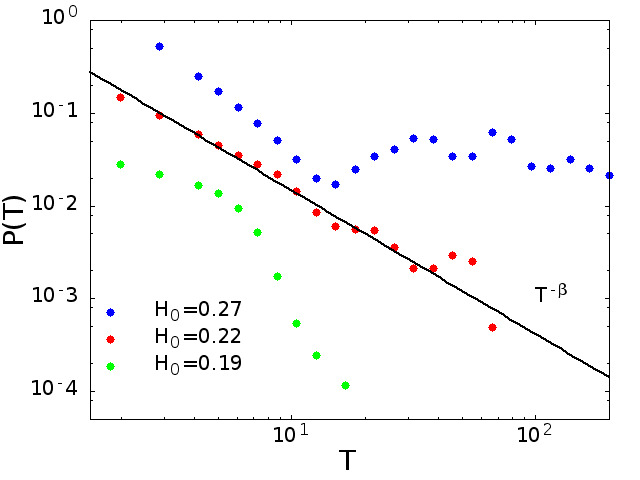}
\\
C \includegraphics[width=5cm]{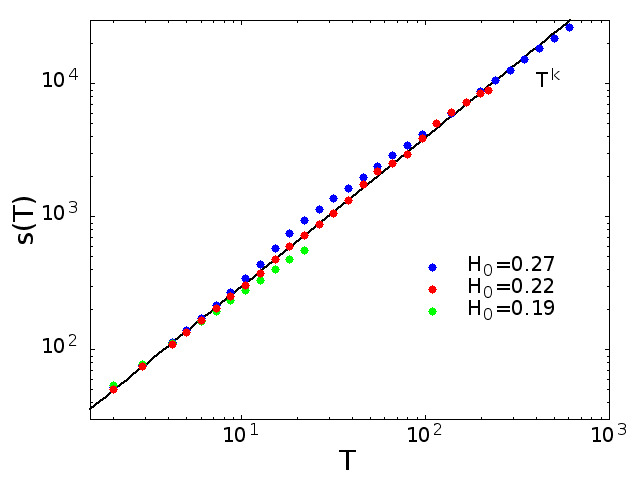}
\end{center}
\caption{(A) Size and (B) duration distribution of the avalanches
  at $N=3000$, $P=2$ and $\alpha=0.06$
   (curves are shifted for clarity).
  For $H_0=0.19$  a subcritical behaviour is observed.
  Power laws are observed near the pseudo-spinodal at $H_0= 22$,
  where the system shows alternation of up and down states.
  Increasing the value of $H_0$ above the pseudo-spinodal,
  the distribution shows a peak signaling a super-critical behaviour.
  The exponents of power laws are $\tau=1.47\pm 0.1$ for the sizes, and $\beta=1.55\pm 0.1$
for the durations.
(C) Average size of the avalanche as a function of the duration. The dependence is
always a power law, with an exponent  $k=1.12\pm 0.01$, in agreement with  Eq.\ (\ref{eq:st})
within errors.
} %
\label{fig-aval2}
\end{figure}

In Figure \ref{fig-aval2}A and B,
 we show the distribution of the sizes and durations of the avalanches 
 for $\alpha=0.06$ and $N=3000$, near the pseudo-spinodal line, $H_0=0.22$, and
 both above and under it.
 We find a clear subcritical behavior at $H_0=0.19$, where the system mostly remains in the down state with very low activity,
a scale-free behavior at $H_0=0.22$ where up/down alternation emerges, and
 a supercritical behaviour with an excess of large avalanches above the pseudo-spinodal line at $H_0=0.27$.
 At $H_0=0.22$, near the pseudo-spinodal line,
 the distributions are well described by power laws,
 with an exponent $\tau=1.47\pm 0.1$ for the sizes and $\beta=1.55\pm 0.1$
 for the durations.

 We used the ``powerlaw'' python package \cite{powerlaw} to
 compute the log-likelihood ratio of the power law fit with respect to an exponential fit,
finding $R=76$ for the size and $R=14$ for the duration (positive values mean that power law is more likely),
 with a significance $p< 10^{-40}$ in both cases,
indicating that the power law fit is much better than the exponential fit.

While the exponent $\tau$ of the sizes is compatible with the largest part of the experimentally measured values, 
the value of $\beta$ found originally (and predicted by models based on a branching process) was $\beta=2$ \cite{Plenz2003}.
However, values similar to the one found here have been observed in some experiments, for example $\beta=1.7\pm 0.2$ in Ref.\ \cite{PRLfriedmann}.

 In Fig.\ \ref{fig-aval2}C, we show the average size of the avalanche as a function
 of its duration, that follows a power law with an exponent
   $k=1.12\pm 0.01$
    which is in agreement, within errors, with the value predicted by the relation
\begin{equation}
k=\frac{\beta-1}{\tau-1}.
\label{eq:st}
\end{equation}

This relation was derived in Ref.\ \cite{sethna2} in relation to crackling noise. 

It can also be derived by this simple reasoning: for values of the duration $T^\prime$ lower than the power law cutoff $T^\ast$, the probability
that an avalanche has a duration $T>T^\prime$ goes as $P(T>T^\prime)\sim (T^\prime)^{1-\beta}$, and analogously $P(s>s^\prime)\sim (s^\prime)^{1-\tau}$
if $s^\prime$ is lower than the cutoff $s^\ast$. Now if $s^\prime$ is the average size of an avalanche of duration $T^\prime$,
and fluctuations in the size fixed the duration can be neglected, then $P(s>s^\prime)\sim P(T>T^\prime)$.
It follows that $(T^\prime)^{k(1-\tau)} \sim (T^\prime)^{1-\beta}$ and
therefore $k$ satisfies Eq. (\ref{eq:st}), at least for sizes and durations below the cut-off.
Notably, the relation $s(T)\sim T^k$
  holds also quite far from critical regime,
  both experimentally
 \cite{PRLfriedmann}
and in our model (see Fig.\ \ref{fig-aval2}C).

Note that the branching process, which is usually connected with the critical behaviour in cortical
networks, predicts values of $\tau=1.5$ and $\beta=2$, so that $k=2$,
substantially greater than the one that we observe. On the other hand, different experiments reported values of the exponent lower than 2,
and more similar to the value that we have measured \cite{naturePhys2015,PRLfriedmann}, with $\tau$ and $\beta$ satisfing the relation (\ref{eq:st}).

A value of $k$ slightly larger than one is in agreement with the fact that avalanches are segments of collective spatio-temporal patterns, having a constant average rate of spikes,
so that the total size of the avalanche is almost proportional to its duration, except for the beginning and end of the burst.
The shape of avalanches in the branching process, on the other hand, corresponds to a rate of spikes having a maximum proportional to the duration $T$ of the avalanche,
giving rise to a total size proportional to $T^2$.

It is interesting that relation (\ref{eq:st}) between the critical  exponents, is verified
in our model and experimentally \cite{naturePhys2015,PRLfriedmann},
while it is not verified in models where power law is not a manifestation of a critical
point \cite{destexhe2}.

\begin{figure}[tbp]
\begin{center}
A \includegraphics[width=5cm]{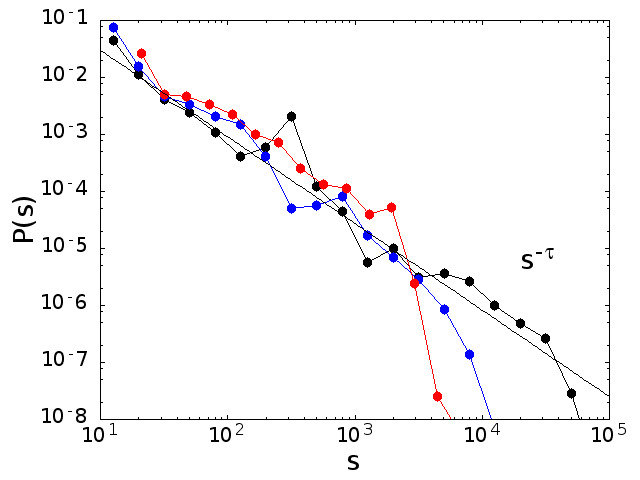}
\\
B \includegraphics[width=5cm]{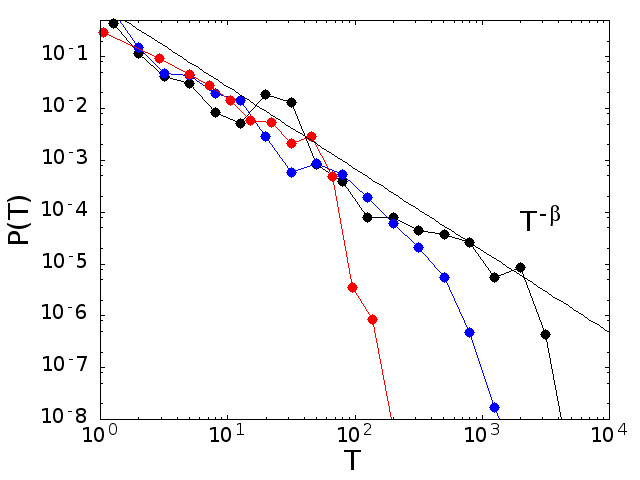}
\\
C \includegraphics[width=5cm]{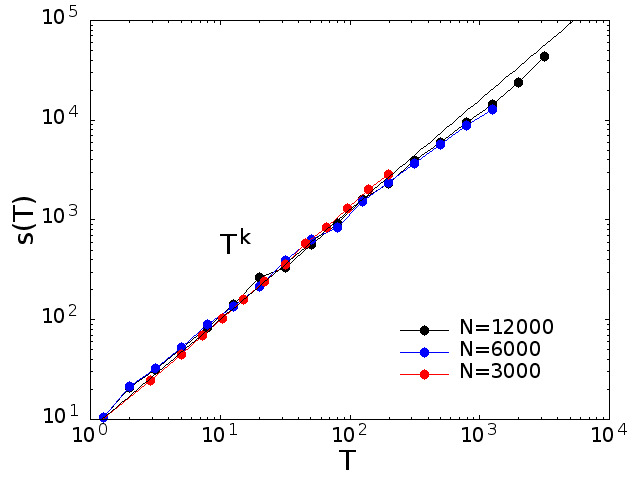}
\end{center}
\caption{
 (A) Size and (B) duration distribution of the avalanches
  at $N=3000$, $6000$ and $12000$, $P=2$, $\alpha=0.06$, and near the
  spinodal instability at respectively $H_0=0.22$, $0.23$ and $0.265$.
For $N=12000$ the exponents are $\tau=1.52\pm 0.05$ for the sizes, and $\beta=1.58\pm 0.05$ for the durations.
(C) Average size of the avalanche as a function of the duration, for the same values of $\alpha$ and $H_0$. For $N=12000$ the exponent of the power law is
$k=1.09\pm 0.05$, in agreement with  Eq.\ (\ref{eq:st}) within errors, taking points with duration $T<50$.
Note that for higher values of the duration the exponent seems to decrease and tend to one,
as observed in \cite{PRLfriedmann}.
}
\label{fig-aval3}
\end{figure}

\begin{figure}[tbp]
\begin{center}
A \includegraphics[width=5cm]{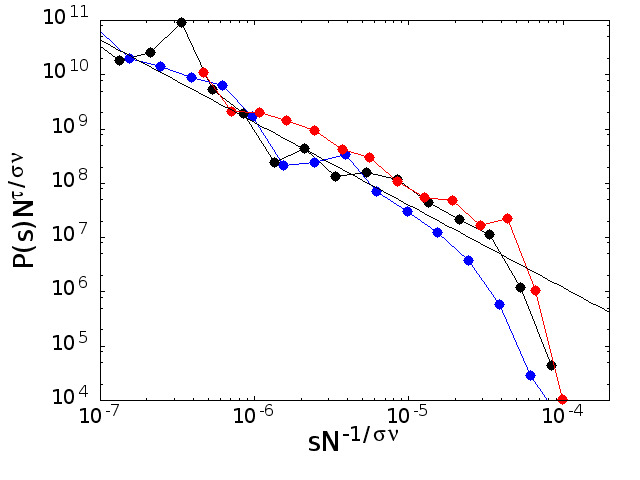}
\\
B \includegraphics[width=5cm]{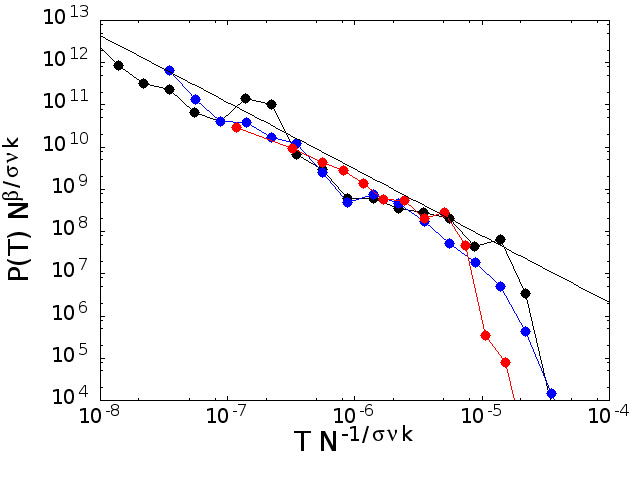}
\end{center}
\caption{Data collapse of the size (A) and duration (B) distributions of the avalanches. The exponent $1/\sigma\nu$ describes the dependence of the cut-off of the sizes
as a function of the system size at the critical point, $s_{\text{max}}\propto N^{1/\sigma\nu}$, while for the durations
$T_{\text{max}}\propto N^{1/\sigma\nu k}$.}
\label{fig-scaling}
\end{figure}

In Fig.\ \ref{fig-aval3}, we show the avalanches distribution at $N=3000$, 6000, 12000, $P=2$, $\alpha=0.06$ and
respectively for $H_0=0.22$, 0.23 and 0.265. The parameters for $N=12000$ are inside the hysteresis region of Fig.\ \ref{fig:phasespace},
where lifetime of the metastable state is longer then experimental time, and near the spinodal instability.
The distributions follow power laws with exponents compatible within errors for different sizes, and with experimental results\cite{plenzlibro,PRLfriedmann}.
For $N=12000$ the exponents found are $\tau=1.52\pm 0.05$ for the sizes, and $\beta=1.58\pm 0.05$ for the durations.
Also in this case we compared the power law fits with the exponential ones, finding a log-likelihood ratio $R=6.2$ for the sizes and $R=7.2$ for the durations,
with a significance $p<10^{-10}$ in both cases.
Notably, as reported in Fig.\ \ref{fig-aval3}A,B, the cut-off of the avalanche distributions scales
with system size, supporting the scale-free behavior of the model near the pseudo-spinodal line.
As shown in Fig.\ \ref{fig-aval3}C,
the average size  as a function of the duration follows a power law also a large sizes,
with an exponent
 $k=1.09\pm 0.05$ at $N=12000$, that is again in agreement within errors with the value
predicted by relation (\ref{eq:st}) and with experimental results \cite{naturePhys2015,PRLfriedmann}.
In Fig.\ \ref{fig-scaling}A and B we show the finite size data collapse of the avalanche size and duration distribution. It is expected that the cutoff of the sizes and durations
scale respectively as $s_{\text{max}}\propto N^{1/\sigma\nu}$ and $T_{\text{max}}\propto N^{1/\sigma\nu k}$ at the critical point, and their distributions are given by
\begin{align*}
P(s)&=N^{-\frac{\tau}{\sigma\nu}}{\tilde P}_s\left(s/s_{\text{max}}\right),
\\
P(t)&=N^{-\frac{\beta}{\sigma\nu k}}{\tilde P}_T\left(T/T_{\text{max}}\right),
\end{align*}
where ${\tilde P}_s(x)$ and ${\tilde P}_s(T)$ are master curves that go as ${\tilde P}_s(x)\propto x^{-\tau}$
and ${\tilde P}_T(x)\propto x^{-\beta}$
at small values of $x$. The best data collapse is given by a value $1/\sigma\nu=2.2$ of the exponent.

Note that, due to the heterogeneity and quenched disorder in the network connectivity,
the region with scale-free avalanches of activity in the model is not limited to a
single point or a single line in the phase-space, but it is an extended region, 
similar to a Griffiths phase \cite{MunozNature}.
A broad region of hysteresis is observed, and at high noise  or small size where there is no hysteresis a broad region of up/down alternation with burst composed of scale-free avalanches is observed.

\section{Discussion}

Scale-free avalanches and critical behaviour in cortical dynamics are
frequently associated with second-order (continuous) phase transitions.
However power law and critical phenomena also
emerge in first-order phase transitions as one enters the metastability
region and approaches the spinodal line, in systems with long range interactions \cite{spinodal3,spinodal3bis}.

Non equilibrium first-order phase transition can be induced by additive noise \cite{jordi,muller}
in spatially extended systems where coupling favors coherent behaviour. Varying the parameters of the systems, or the noise,
the order of the phase transition  may change.  First-order phase transition with a coexistence
region where the system displays hysteresis, and a
crossover to a second-order transition for large values of the
noise, has been studied in a variety of systems such as surface growth \cite{giada2000}.
Hysteresis in a stochastic non-leaky integrate and fire model has been studied in \cite{kaltenbrunner}, 
but scale free avalanches were not investigated.
A pioneer model that has hypothesized a new scenario for cortical dynamics, combining self organized criticality with a first order transition, is the one 
studied in Refs.\ \cite{levinaPRL2009,levinanatphys,geiselreview}.
More recently \cite{munozPRL2016} it has been suggested that cortical networks are not self-organized to a critical point (SOC),
as usually considered, but to a region of bistability (SOB)  near a first order transition.
A most referenced model that displays criticality in a network of leaky integrate-and-fire neurons, has been studied 
in Ref.\ \cite{millman}. In that model, however, criticality emerges only with a definition of avalanches that takes in account the causality of different firings.
If one uses a criterion based only on temporal binning and proximity, as done usually in experiments where causality is not observable,
one finds exponential distributions of avalanches, and no critical behaviour is observed \cite{maritan}.

In this paper, we studied a model of leaky integrate and fire neurons, characterized by structured long range connectivity, corresponding to the encoding of spatio-temporal patterns.
We showed that the model exhibits a first order transition between a down state characterized by low activity, and an up state characterized by the collective replay of one of the
spatio-temporal patterns encoded in the network.
Notably the role of noise is crucial. Indeed,
depending on the noise and size of the system,
one observes hysteresis (low noise or large size) or up/down alternation (high noise or small size) at the transition.

Increasing the size of the system, the lifetimes of states increase, and one observes hysteresis also for values of the noise that showed alternation of states at smaller sizes,
showing that the alternating behaviour is a finite size effect, and the underlying transition is of a first order kind.
Both in the region of hysteresis, approaching the spinodal instability, and in the region of alternation, 
we observe scale-free bursts of activity (avalanches). Notably this was found identifying avalanches using the same criterion of temporal proximity used in experiments.

While scale-free avalanches alone are not sufficient to assess criticality \cite{destexhe1,destexhe2,beggs1}, we have independently identified a (first-order) transition by the 
discontinuity in the rate and the hysteretic behaviour. Power law distributions, and a peak in the normalized variance,
are then observed near the edge of the spinodal instability, as is expected for a first order
transition in a model with long range connections.

The model therefore incorporates both criticality and the functioning of the network as a memory, or reservoir of dynamical patterns.
When the system is posed at the edge of the instability, it shows spontaneous ongoing activity with critical scale free behavior.
However the state is a metastable one. If a cue stimulation (a short train of input spikes,
with order similar to one of the stored patterns) is given, then the
system switches in the persistent up state and responds with a (non-critical) sustained replay of the pattern stimulated.
This behaviour reminds the one observed experimentally in Ref.\ \cite{naturePhys2015},
where a transient state characterized by large non-critical avalanches is observed in response to an external stimulus.

The exponents of the size and duration distribution, 
and the exponent $k$ giving the dependence of the size on the duration of the avalanche, $s(T)\sim T^k$
are compatible with the range of values found experimentally \cite{plenzlibro,naturePhys2015,PRLfriedmann}.
The value of $k$ near to one is due to the mechanism of avalanche propagation. Indeed
in our case avalanches are segments of patterns having an almost constant average
spiking rate, so that the total size is almost proportional to the duration.

  In a branching process model \cite{munozPRE2017} it
  was shown that exponents of size and duration distributions are not universal, but they vary depending on a small external driving of the system.
The effect of the driving, in the class of branching processes that they consider,
is to merge smaller avalanches to form larger ones, therefore the relative weight of larger avalanches increases, and exponents decrease with the driving,
while the exponent $k$ remains equal to 2 independently from the driving.
In our model, on the other hand, avalanches are related to the emergence of a collective coordinated activity. Therefore
the effect of noise is not only to merge avalanches, but also to hinder their propagation, decreasing the probability of longer avalanches.
A higher noise decreases the lifetime of the
metastable states, and hysteresis turns into up/down alternation, and even higher noise makes the region of up/down alternation broader.

Another paper that has considered the effect of the noise is Ref.\ \cite{garcia}. They studied a ``cortical branching model'', that has a non equilibrium phase
transition only in the limit of zero spontaneous activation (that has a similar role of our noise), and a quasi-critical behaviour on the Widom line at finite values of the
spontaneous activation, with a broadening of susceptibility. Also in our case we observe a broadening of the susceptibility (normalized variance), with the increase of the noise.
However, in our model we observe a (first order) transition also at non zero values of the noise, that produces (for not too low noise) an alternation between
up and down states. Therefore the broadening of the variance is not connected to a Widom line,
but to the broadening of the region where alternation of up and down states is observed.

The main characteristic of our model is the structure of the connections, that are not chosen randomly, but are the result of a learning rule inspired by spike-time dependent
plasticity (STDP), where different
spatio-temporal patterns (corresponding to different sequences of firing of the neurons) are encoded.
Connections are set at the beginning, and are held fixed during the dynamics of the network.
Due to the fact that the learning kernel has a zero integral over time (see Appendix), connections are characterized by a balance between excitation and
inhibition, which is one of the ingredients to observe a critical behaviour, as observed experimentally \cite{shew-2,mazzoni} and also in models \cite{lucillachaos2017}.
However, balance is not the only ingredient,
since the topology and structure of the connectivity, with collective patterns carved as attractors of the dynamics, are crucial to observe the non-equilibrium first-order transition.
Preliminary results indeed indicate that, reshuffling the connections randomly between neurons, the transition disappears.
One observes on the contrary a continuous increasing of the spiking rate when the strength of the connections is increased, with a normalized variance always near to one,
showing that the dynamics is Poissonian, and no critical behaviour is observed \cite{wirn16}.
This is also in agreement with recent results, showing that topology is crucial for the emergence of critical states \cite{massobrio}.

The presence of a non-equilibrium first order transition and the critical precursor phenomena in our model are crucially related
to the interplay between noise and a connectivity which promotes collectivity.
Criticality emerges naturally %
near the edge of an instability,
in an associative memory network, with many metastable dynamical states.

Another model that study criticality together with associative memory was proposed in \cite{levinaasso}.
In their model, a Hebbian learning rule is used to store static patterns.
However, they found that Hebbian learning alone destroys criticality even when the synaptic strength is properly scaled.
Applying an optimization procedure that drives the synaptic couplings either toward the critical regime, or toward the memory state in an alternating fashion,
they finally arrive at a configuration both critical and that retains an associative memory.
The reason why in our model the learning procedure does not destroy criticality may be due to
the difference in the learning rule, that in our case is based on STDP and stores dynamical attractors, as opposed to static ones.

Previous studies based on the branching process have explained the
repeatability of spatiotemporal patterns \cite{beggs2,beggs3}, together with power laws in avalanches distribution.
In their model however, patterns are not shown to
be attractors of the dynamics in any parameter space region.
They show repeatability only in the critical region, where the dynamics is ``neutral'' (Lyapunov exponent equal to zero).
On the other hand, in our model the stored
spatiotemporal patterns emerge as collective attractors of the
dynamics in the region above the transition.
The systems is therefore both able to work as a stimulus-activated
reservoir of spatiotemporal attractors, and as a more flexible device
when used at the border of the instability.

In our model, an alternation of up and down states is obtained with a fixed value of the excitability $H_0$,
inside a certain range of external parameters near the transition.
The same value of $H_0$, lowering the noise, gives rise to hysteresis, with persistence of one or the other state, depending on the previous history.
However, it is plausible that the brain is able to change its state also by changing the value of $H_0$, going out of the critical region
toward a persistent up (more suited for either spontaneous or cue-triggered reactivation of previous experience)
or down state (which favors faithful representation of sensory inputs) depending on the different behavioural state.
For example Ref.\ \cite{shew2015} shows that focused attention pulls the system out of criticality towards subcriticality.
The switch between different states, between sleep and wakefulness or from inattentive to vigilant states,
may be induced by specific neuromodulators that, among other effects, can also change the efficacy of the connections.
Neuromodulation is important for regulating brain states \cite{lee2012}, but the specific mechanisms of these switching are not yet well understood.

Another important ingredient of the network connectivity in our model is the presence of a small percentage of neurons that has higher incoming connection strengths, and is responsible
for focusing the noise and initiating the collective activity (and avalanche propagation).
The presence of a few highly active sites, driving cortical neural activity (leaders), has been reported experimentally \cite{orlandi,pasquale2017,plos18,eckmann}.
Notably it has been shown that these leader sites are reliably and rapidly recruited within both spontaneous and evoked bursts \cite{pasquale2017}.
As shown in Ref.\ \cite{orlandi}, initiation of
bursts of collective activity in cultured networks is a noise-driven nucleation phenomenon.
The nucleation sites seem to be highly localized, they collect and amplify activity originated elsewhere.
This noise focusing effect is realized in the model with this higher $H_0$ to incoming connections to a bunch of
neurons which focus noise and cooperate to initiate the emergence of the pattern.

There are some predictions that could be investigated in experiments, to discriminate between the first-order transition scenario considered here and other models.
A prediction is that, lowering the noise, the lifetime of the states increases,
and the system goes from a phase with alternation of up and down states, to a phase characterized by metastability and hysteresis.
The noise can be related for example to spontaneous neuro-transmitter release.
Another prediction of our model is that, increasing the strength of the connections but mantaining the balance between excitation and inhibition,
the patterns that in the critical region appear during the alternation of up and down states, become more attractive, and can be replayed for a longer time.
To our knowledge this kind of experiment has not been realized. What has been done is something quite different, that is changing the balance between
excitation and inhibition.
This tunes the network into a phase characterized by high activity, far from the critical regime and with an excess of long avalanches.
It is not clear however if this corresponds to the same kind of transition that we observe. 

To our knowledge, this is the first leaky integrate and fire model, which shows how both dynamical attractors and neuronal avalanches converge in a single cortical model.
This work therefore may help to link the bridge between criticality and the need to have a reservoir of spatiotemporal metastable memories.

\section*{Appendix: The model}

We simulate a network of $N$ spiking neurons, modelled as leaky integrate-and-fire units
and represented by the Spike Response Model \cite{gerstner},
in presence of a Poissonian noise distribution. We study the spontaneous dynamics of the neurons connected by a sparse structured connectivity in absence of any external inputs. 
Between consecutive spikes, the membrane potential of neuron $i$ is given by
\begin{align}
u_i (t) &=\sum_{j}\sum_{t_i<t_j<t}J_{ij}\left[e^{-(t-t_j)/\tau_m}-e^{-(t-t_j)/\tau_s}\right]
\nonumber\\
&+\sum_{t_i<\hat{t}_i<t}J({\hat{t}_i})\left[e^{-(t-\hat{t}_i)/\tau_m}-e^{-(t-\hat{t}_i)/\tau_s}\right],
\label{eqSRM}
\end{align}
where $J_{ij}$ is the synaptic strength between presynaptic neuron $j$ and postsynaptic neuron $i$, $t_j$ are the spiking times of neuron $j$ coming after the last spike $t_i$ of neuron $i$,
$\hat{t}_i$ are random times extracted from a Poissonian distribution with rate $\rho=1$ ms$^{-1}$, $J({\hat{t}_i})$ is a Gaussian variable extracted at time $\hat{t}_i$
with zero mean and standard deviation
$\sqrt{\frac{\alpha N}{3000\rho}\sum_{j}J_{ij}^2}$,
$\tau_m$ is the characteristic time of membrane ($\tau_m=10$ ms) and $\tau_s$ is the characteristic time of synapse ($\tau_s=5$ ms).
When the membrane potential $u_i(t)$ hits the threshold $\Theta=1$, it is reset to zero, and spikes are transmitted to all the neurons that receive input from neuron $i$.

The strengths of the connections are determined by
a learning rule \cite{EPL,JCN,plos,frontiers}, inspired by STDP (spike-time dependent plasticity), which gives rise to a highly heterogeneous and disordered distribution of weights. 

We build the connections $J_{ij}$ forcing the network to store $P$ spatio-temporal patterns.
Each pattern is a periodic train of spikes, with one spike per neuron and per cycle, with the neuron $i$ firing at times $t_i^\mu+nT$, with $t_i^\mu$ randomly and uniformly extracted
in the interval $[0,T^\mu]$.
In the present work, we use a number of neurons
 between $N=3000$ and $N=12000$,
and a number of patterns between $P=2$ and $P=10$, with period $T^{\mu}=333$ ms.
After the learning stage, the strength of connection $J_{ij}$ is given by
\begin{equation}
  {J}_{ij}=\frac{f_iH_0}{N}\sum_{\mu=1}^P\sum_{n=-\infty}^{\infty}A(t_i^{\mu}-t_j^{\mu}+nT^{\mu})
  \label{eq:J}
\end{equation}
where $A(\tau)$ is the STDP learning window \cite{bipoo,markram}, given by 
\begin{equation}
A(\tau)=
\left\{\begin{array}{ll}
a_p e^{-\tau /T_p}-a_D e^{-\eta\tau /T_p} & \mbox{if $\tau > 0$},
\\
a_p e^{\eta \tau /T_D}-a_D e^{\tau/T_D}   & \mbox{if $\tau < 0$},
\end{array}\right.
\end{equation}
with $a_p=A_0/[1 + \eta T_p/T_D]$, $a_D=A_0/[\eta + T_p/T_D]$, $A_0=3000$, $T_p = 10.2$ ms, $T_D = 28.6$ ms, and $\eta = 4$.
To take account of the heterogeneity of the neurons, we use two values of $f_i$, $f_i=1$ for "normal" neurons and $f_i=3$ for "leader" neurons,
i.e. neurons with higher incoming connection strengths, that amplify activity initiated by noise \cite{orlandi,pasquale2017,plos18,eckmann}.
In other words, leaders are neurons that fire more than other ones and give rise to a cue able to initiate the short collective replay.
They are chosen as a fraction of 3\% of neurons with consecutive phases, for each pattern $\mu$.
The connection  $J_{ij}$  between neurons $i$ and $j$ does not depend therefore on the spatial distance between them,
if they are embedded in a 2D or 3D space. Therefore
this form of the connections is a ``long range'' one, for which one could expect a ``mean field like'' behaviour, with long lifetimes (infinite in the thermodynamic limit)
of the metastable states. Long lifetimes can be expected also if connections are not independent from the distance, but the range is not too small.

Note that $J_{ij}$ are proportional to $N^{-1}$, so that the noise is independent from $N$ at fixed value of $\alpha$.
On the other hand, due to the shape of the STDP learning kernel that has time integral equal to zero,
this learning procedure assures the balance between excitation and inhibition, i.e. $\sum_j J_{ij}$ is of order $1/\sqrt{N}$.
  At the end of the learning procedure, part of the connections are positive (excitatory) and part are negative (inhibitory).
  Inhibitory neurons are not explicitly simulated, but negative
  negative connections can be considered  as connections mediated by fast inhibitory interneurons.
Alternatively, one could introduce a global inhibition and explicitly simulate only the positive connections.

\begin{figure}[tbp]
\begin{center}
\includegraphics[width=5cm]{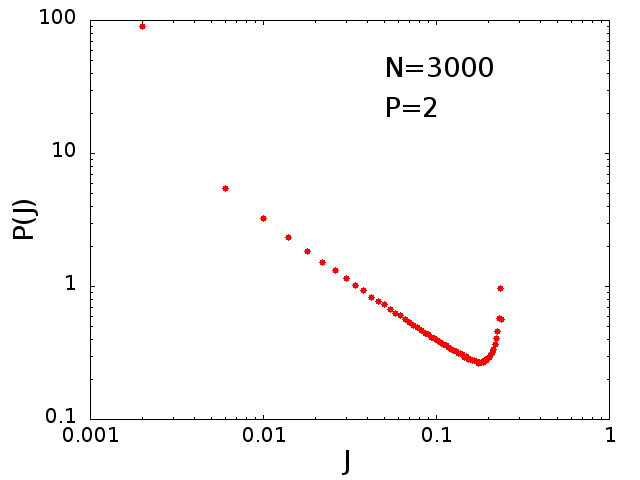}
\end{center}
\caption{Distribution of the positive connections after the learning procedure, for $N=3000$ neurons and $P=2$ patterns.}
\label{fig-weights}
\end{figure}

The result of learning multiple spatio-temporal patterns, each with quenched randomly-chosen phase ordering,
gives rise to quenched disorder.
The distribution of weight that results from this learning procedure is highly heterogeneous, with many small connections and few strong ones.
In Fig.\ \ref{fig-weights} we show the distribution of the positive weights for $N=3000$ neurons and $P=2$ patterns. 
The distribution is very skewed and long-tailed, as observed in the cortex \cite{sjostrom,loewenstein}
and in other STDP-based models \cite{yassin,effenberger,tosi}.
Note however that in our model the distribution of the weights is not a sufficient condition to determine the dynamical phase transition. Indeed, shuffling the connections
leaving their distribution unchanged, this kind of transition disappears \cite{wirn16}. It seems therefore that the topology of the network, such as the
relative abundance of motifs, is crucial for the manifestation of the first order dynamical transition.

To get a sparse connectivity, like in the brain cortex, we prune the smallest connections.
The pruning procedure keeps still the balance between excitation and inhibition, and leaves only $30\%$ of the original connections.
Once this connectivity structure is built, it is kept fixed during all the network dynamics simulations.

Note that, apart from the quenched random values of the times $t_i^{\mu}$ defining the encoded patterns, and $f_i$ defining the leader neurons,
the dynamics of the model depends only on the parameters $\alpha$, determining the strength of the noise, and $H_0$, determining the strength of the connections.

%

\end{document}